\documentclass[12pt]{article}
\usepackage{epsf}
\usepackage{a4wide,graphicx}
\usepackage{cite}
\usepackage{subfigure}

\title{Study of the $J/\psi \to \phi (\omega) f_2(1270)$,
$J/\psi \to \phi (\omega) f'_2(1525)$ and $J/\psi \to K^{*0}(892) \bar{K}^{*\,0}_2(1430)$
decays}

\author{
A. Mart\'inez Torres$^{1,4}$, L.S. Geng$^1$, L.R. Dai$^{2,4}$, Bao Xi Sun$^{3,4}$, E.~Oset$^{1,4}$ \\ and B.S. Zou$^4$\\
{\small{\it $^1$Departamento de F\'{\i}sica Te\'orica and IFIC,
Centro Mixto Universidad de Valencia-CSIC,}}\\
{\small{\it Institutos de
Investigaci\'on de Paterna, Aptdo. 22085, 46071 Valencia, Spain.}}\\
{\small{\it $^2$ Department of Physics, Liaoning Normal University, Dalian, 116029, China.}}\\
{\small{\it $^3$ Institute of Theoretical Physics, College of Applied Sciences,
Beijing University}}\\
{\small{\it of Technology, Beijing 100124, China.}}\\
{\small{\it $^4$ Institute of High Energy Physics, and Theoretical Physics Center}}\\
{\small{\it for Science Facilities, CAS, Beijing 100049, China.}}
}
\date{\today}
\begin{document}
\maketitle

\begin{abstract}
We present an approach to study the decay modes of the  $J/\psi$ into a vector
meson and a  tensor meson, taking into account the nature of the $f_2(1270)$,
$f'_2(1525)$, $\bar{K}^{*\,0}_2(1430)$ resonances as dynamically generated
states from the vector meson-vector meson interaction. We evaluate four ratios of partial
decay widths in terms of a flavor dependent OZI breaking parameter and the
results obtained compare favorably with experiment, although the experimental uncertainties are still large. Further refinements of the data would provide a more stringent test on the theoretical approach. The fit to the data is possible due to the particular strength and sign of the couplings of the resonances to pairs of vector mesons given by the theory.
\end{abstract}

\section{Introduction}
The decay of $J/\psi$ into a vector meson $\phi$ or  $\omega$ and a scalar
meson $\sigma(600)$ or $f_0(980)$ was used in
\cite{ulfjose,palochiang,lahde,liu} as a test of chiral dynamics and the
dynamically generated character of the low lying scalar mesons
$\sigma(600)$ and $f_0(980)$.  The process proceeds through an OZI violating
strong interaction, which makes complicated
its quantitative microscopical study. On the other hand the reaction offers a
genuine simplifying factor since the  $J/\psi$, a $c \bar{c}$ state, qualifies as a
 singlet of SU(3) and then one can relate the $J/\psi$ decays into
 $\phi$ or  $\omega$ and the different scalar mesons,  up to a global
 normalization factor, which would entail the difficulties of a microscopic evaluation.
 The use of chiral dynamics, assuming the low energy scalars as dynamically generated mesons
  \cite{liu,npa,ramonet,kaiser,markushin,rios,arriola,juanlarge}, allowed a good representation of the experimental data for this set of $J/\psi$ decays.  Analogous to the
 decay modes mentioned above
are the modes $J/\psi \to \phi (\omega) f_2(1270)$,
 $J/\psi \to \phi (\omega) f'_2(1525)$ and $J/\psi \to K^{*0}(892) \bar{K}^{*\,0}_2(1430)$. The
 novelty is that one is replacing the scalar states by the tensor resonances
$ f_2(1270),f'_2(1525), \bar{K}^{*\,0}_2(1430) $.  Recent steps
generating dynamically some mesons from the interaction of pairs of
vector mesons allow us to retake the ideas of
\cite{ulfjose,palochiang,lahde,liu}  and extend them to the $J/\psi$
decay into a vector meson and this new family of tensors.  Indeed,
in \cite{raquel} it was shown that the  $f_2(1270)$ and  $f_0(1370)$
states appear naturally as bound states of $\rho \rho$ using the
interaction provided by the hidden gauge Lagrangians
\cite{hidden1,hidden2,hidden3}. An extension to SU(3) of the former
work
 \cite{gengvec},
studying the interaction of pairs of vectors, shows that there are 11 states
dynamically generated, some of which can clearly be associated to known
resonances.  The $ f_2(1270),f'_2(1525), \bar{K}^{*\,0}_2(1430) $ resonances are some of
those appearing clearly in that latter work. In view of this, one can take the
vector-vector components of these states, provided in terms of the couplings of
the resonances to the different vector-vector states, and using similar
arguments as done in \cite{ulfjose,palochiang,lahde,liu} one can make predictions
for the ratios of the partial decay widths into all these channels. The approach
turns out to be rewarding. In terms of one necessary parameter appearing in the
approach, one can make predictions for the ratios of all these partial decay
widths, which are in agreement with data.

\section{Formalism for $J/\psi\rightarrow \mathrm{VT}$}
Following the usual approach to deal with dynamically generated resonances one has to couple the external sources to the components of these states,
which then interact among themselves to provide the resonances.
In this case, the primary study has to be $J/\psi\rightarrow V{V'V'}$, where
$V'V'$ are the pairs of vector mesons that lead to the desired resonances upon interaction.

\begin{figure*}[h!]
\centerline{\includegraphics[scale=0.5]{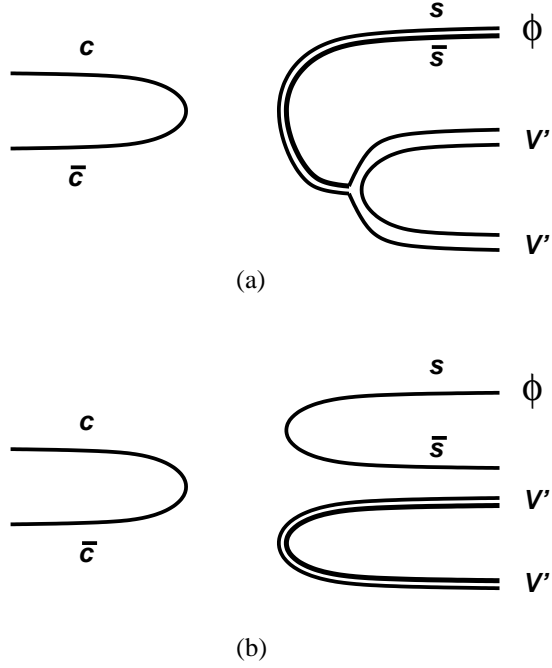}}
\caption{Two different OZI forbidden mechanisms of $J/\psi\rightarrow\phi V'V'$.}\label{ccbar}
\end{figure*}

Following Ref.\cite{ulfjose} we depict in Fig.\ref{ccbar} the mechanisms violating OZI that lead to the
primary step $J/\psi\rightarrow V{V'V'}$. The sum of these two mechanisms indicate
that the lower $q\bar{q}$ pair in the figure that hadronizes into
$V'V'$ can be either $s\bar{s}$ or $u\bar{u}+d\bar{d}$ (to respect isospin symmetry). In
Ref.~\cite{ulfjose} this was taken into account empirically in terms of a $\lambda_\phi$ parameter, such that the combination in the lower part of the diagram was given by
\begin{equation}
 s\bar{s}+\lambda_\phi\frac{1}{\sqrt{2}}(u\bar{u}+d\bar{d})
\end{equation}
up to a global normalization factor. Further studies in Ref.~\cite{ulfjose}
allowed to relate the $J/\psi\rightarrow\omega\pi\pi$ decay amplitude to the $\lambda_\phi$ parameter.
In Ref.~\cite{palochiang} a group theoretical approach was used which
was equivalent to the one of Ref.~\cite{ulfjose}. We follow here the approach of Ref.~\cite{palochiang}.

\begin{figure*}[h!]
\centerline{\includegraphics[scale=0.5]{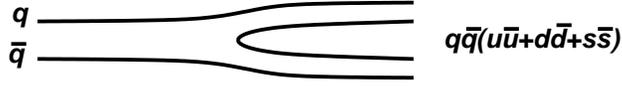}}
\caption{A schematic representation of $q\bar{q}\rightarrow q\bar{q}(u\bar{u}+d\bar{d}+s\bar{s})$.}
\label{ccbar2}
\end{figure*}
By taking $\omega$, $\phi$ to be ideal mixtures of singlet and octet SU(3) states we have
\begin{eqnarray}
 \phi&=&s\bar{s}=\sqrt{\frac{1}{3}}V_1-\sqrt{\frac{2}{3}}V_8\nonumber\\
\omega&=&\frac{1}{\sqrt{2}}(u\bar{u}+d\bar{d})=\sqrt{\frac{2}{3}}V_1+\sqrt{\frac{1}{3}}V_8
\end{eqnarray}
In this sense, the $J/\psi\rightarrow\phi s\bar{s}\rightarrow \phi V'V'$ and
$J/\psi\rightarrow\phi\frac{1}{\sqrt{2}}(u\bar{u}+d\bar{d})\rightarrow\phi V'V'$ amplitudes
will involve matrix elements of
\begin{equation}
 \left\langle\Bigg(\sqrt{\frac{1}{3}}V_1-\sqrt{\frac{2}{3}}V_8\Bigg)\Bigg(\sqrt{\frac{1}{3}}V_1-\sqrt{\frac{2}{3}}V_8 \Bigg)\Big|T\Big|0\right\rangle
\end{equation}
and
\begin{equation}
 \left\langle\Bigg(\sqrt{\frac{1}{3}}V_1-\sqrt{\frac{2}{3}}V_8\Bigg)\Bigg( \sqrt{\frac{2}{3}}V_1+\sqrt{\frac{1}{3}}V_8\Bigg)\Big|T\Big|0\right\rangle
\end{equation}
respectively, where $|0\rangle$ signifies the SU(3) singlet corresponding to the $J/\psi$ state. Taking into account that $T$ is SU(3) invariant, we obtain the amplitudes
\begin{equation}\label{eq:phiamp}
 \frac{1}{3}T^{(1,1)}+\frac{2}{3}T^{(8,8)}\quad\quad\mbox{and}\quad\quad
\frac{\sqrt{2}}{3}T^{(1,1)}-\frac{\sqrt{2}}{3}T^{(8,8)}
\end{equation}
for the two possibilities. Similarly the $J/\psi\rightarrow\omega s\bar{s}\rightarrow\omega V'V'$
and $J/\psi\rightarrow\omega\frac{1}{\sqrt{2}}(u\bar{u}+d\bar{d})\rightarrow\omega V'V'$ would
have the amplitudes
\begin{equation}
 \left\langle\Bigg(\sqrt{\frac{2}{3}}V_1+\sqrt{\frac{1}{3}}V_8\Bigg)\Bigg(\sqrt{\frac{1}{3}}V_1-\sqrt{\frac{2}{3}}V_8 \Bigg)\Big|T\Big|0\right\rangle
\end{equation}
\begin{equation}
 \left\langle\Bigg(\sqrt{\frac{2}{3}}V_1+\sqrt{\frac{1}{3}}V_8\Bigg)\Bigg(\sqrt{\frac{2}{3}}V_1+\sqrt{\frac{1}{3}}V_8 \Bigg)\Big|T\Big|0\right\rangle
\end{equation}
which in terms of $T^{(1,1)}$ and $T^{(8,8)}$
read as
\begin{equation}\label{eq:omegaamp}
 \frac{\sqrt{2}}{3}T^{(1,1)}-\frac{\sqrt{2}}{3}T^{(8,8)}\quad\quad
\mbox{and}\quad\quad
\frac{2}{3}T^{(1,1)}+\frac{1}{3}T^{(8,8)}
\end{equation}
respectively.

Since we will not be interested in the absolute normalization, we can work
with the ratio of amplitudes
\begin{equation}\label{eq:nu}
 \nu=\frac{T^{(1,1)}}{T^{(8,8)}}
\end{equation}
and the amplitudes of Eqs.~(\ref{eq:phiamp},\ref{eq:omegaamp}) are given in Table \ref{coef}.
\begin{table}[h!]
\caption{Transition amplitudes of $J/\psi\to\phi(\omega)$ and extra components $s\bar{s}$, $\frac{1}{\sqrt{2}}(u\bar{u}+d\bar{d})$ that hadronize later into $V^\prime V^\prime$. Also the transition amplitude for $J/\psi\to K^{*\,0}$ and $s\bar{d}$ that hadronizes latter.}\label{coef}
\vspace{0.5cm}
\centering
\begin{tabular}{c|c|c|c}
\hline\hline
&$\phi$ production&$\omega$ production&$K^{*\,0}$ production\\
\hline
&&&\\
lower $s\bar{s}$&$\frac{1}{3}\nu+\frac{2}{3}$&$\frac{\sqrt{2}}{3}\nu-\frac{\sqrt{2}}{3}$&\\
&&&\\
\hline
&&&\\
lower $\frac{1}{\sqrt{2}}(u\bar{u}+d\bar{d})$&$\frac{\sqrt{2}}{3}\nu-\frac{\sqrt{2}}{3}$&$\frac{2}{3}\nu+
\frac{1}{3}$&\\
&&&\\
\hline
&&&\\
lower $s\bar{d}$&&&1\\
\end{tabular}
\end{table}

The case of $K^*(892)$ production is easier since it has only an SU(3) octet component.
Hence, the amplitude for $J/\psi\rightarrow K^*(892) q\bar{q}\,\,$ is $T^{(8,8)}$,
or simply unity after the normalization of Eq.~(\ref{eq:nu}).

The next step is to see how the lower $q\bar{q}$ components in the diagrams of Fig. \ref{ccbar}
hadronize into $V'V'$, as depicted in Fig. \ref{ccbar2}. For this we follow the quark line counting. The usual SU(3) meson matrix in terms of $q\bar{q}$ is given by
\begin{equation}
 M=\left(\begin{array}{ccc}u\bar{u}&u\bar{d}&u\bar{s}\\
                           d\bar{u}&d\bar{d}&d\bar{s}\\
                           s\bar{u}&s\bar{d}&s\bar{s}
\end{array}\right)
\end{equation}
with the property that
\begin{equation}
 M\cdot M=M\times(u\bar{u}+d\bar{d}+s\bar{s}).
\end{equation}
This symbolizes the hadronization process implicit in the diagram of Fig. \ref{ccbar2}.
We can write the meson matrix $M$ in terms of the vector states
\begin{equation}
 \textbf{V}=\left(\begin{array}{ccc}
          \frac{1}{\sqrt{2}}\rho^0+\frac{1}{\sqrt{2}}\omega&\rho^+&K^{*+}\\
           \rho^-&-\frac{1}{\sqrt{2}}\rho^0+\frac{1}{\sqrt{2}}\omega&K^{*0}\\
           K^{*-}&\bar{K}^{*0}&\phi
         \end{array}
\right)
\end{equation}
such that $\textbf{V}\cdot \textbf{V}$ will be equivalent to $M\cdot M$
\begin{equation}
 \textbf{V}\cdot \textbf{V}\equiv M\times(u\bar{u}+d\bar{d}+s\bar{s})
\end{equation}
and hence, the matrix elements of $M$ prior to hadronization can be
associated to $\textbf{V}\cdot \textbf{V}$, which will give us the weight of the different
vector-vector components after hadronization.
Hence, we associate
\begin{equation}
 s\bar{s}\rightarrow K^{*-}K^{*+}+\bar{K}^{*0}K^{*0}+\phi\phi,
\end{equation}
\begin{equation}
 \frac{1}{\sqrt{2}}(u\bar{u}+d\bar{d})\rightarrow
\frac{1}{\sqrt{2}}(\rho^0\rho^0+\rho^+\rho^-+\rho^-\rho^+ +\omega\omega+K^{*+}K^{*-}+K^{*0}\bar{K}^{*0}).
\end{equation}
In order to project these combinations onto the physical states of
$\rho\rho$, $K^*\bar{K}^*$, $\omega\omega$, $\phi\phi$ with isospin
0, we recall the normalization and phases used in
Ref.~\cite{gengvec}, $\rho^+=-|1,+1\rangle$,
$K^{*-}=-|1/2,-1/2\rangle$ and the unitary normalization (an extra
factor $\frac{1}{\sqrt{2}}$ in the case of identical particles, or
symmetrized ones, to ensure the resolution of identity in the sum
over intermediate states):
\begin{eqnarray}
 |\rho\rho,I=0\rangle&=&-\frac{1}{\sqrt{6}}|\rho^0\rho^0+\rho^+\rho^-+\rho^-\rho^+\rangle\nonumber\\
 |K^*\bar{K}^*,I=0\rangle&=&-\frac{1}{2\sqrt{2}}|K^{*+}K^{*-}+K^{*-}K^{*+}+K^{*0}\bar{K}^{*0}+\bar{K}^{*0}K^{*0}\rangle\nonumber\\
|\omega\omega,I=0\rangle&=&\frac{1}{\sqrt{2}}|\omega\omega\rangle\nonumber\\
|\phi\phi,I=0\rangle&=&\frac{1}{\sqrt{2}}|\phi\phi\rangle
\end{eqnarray}
Thus, the hadronized $s\bar{s}$, $\frac{1}{\sqrt{2}}(u\bar{u}+d\bar{d})$ components
lead to the weights of Table \ref{coef2}.
\begin{table}[h!]
\caption{Weights of $s\bar{s}$ and $\frac{1}{\sqrt{2}}(u\bar{u}+d\bar{d})$ into the different $I=0$ $V^\prime V^\prime$ components.}\label{coef2}
\vspace{0.3cm}
\centering
\begin{tabular}{c|cccc}
\hline\hline
&$\rho\rho$&$K^*\bar{K}^*$&$\omega\omega$&$\phi\phi$\\
\hline

lower $s\bar{s}$&0&-$\frac{1}{\sqrt{2}}$&0&$\frac{1}{\sqrt{2}}$\\

lower $\frac{1}{\sqrt{2}}(u\bar{u}+d\bar{d})$&-$\frac{\sqrt{3}}{2}$&-$\frac{1}{2}$&$\frac{1}{2}$&0
\end{tabular}
\end{table}

Considering the factors of Table \ref{coef2} and the amplitudes of Table \ref{coef} for
$\phi$ and $\omega$ production together with the hadronization of $s\bar{s}$ and $\frac{1}{\sqrt{2}}(u\bar{u}+d\bar{d})$ components, we obtain the weights $W_i$ for
$J/\psi\rightarrow\phi(\omega)V'V'$ (Table \ref{weight2}).
\begin{table}[h!]
\caption{Weights for the processes $J/\psi\rightarrow\phi(\omega)V'V'$ with $V^\prime V^\prime$ in isospin $I=0$. }\label{weight2}
\vspace{0.3cm}
\centering
\begin{tabular}{c|cccc}
\hline\hline\\
&$\rho\rho$&$K^*\bar{K}^*$&$\omega\omega$&$\phi\phi$\\
\hline
&&&&\\
$W^{(\phi)}$&$-\frac{1}{\sqrt{6}}(\nu-1)$&$-\frac{\sqrt{2}}{6}(2\nu+1)$&$\frac{1}{3\sqrt{2}}(\nu-1)$&$\frac{1}{3\sqrt{2}}(\nu+2)$\\
&&&&\\
$W^{(\omega)}$&$-\frac{1}{2\sqrt{3}}(2\nu+1)$&$-\frac{1}{6}(4\nu-1)$&$\frac{1}{6}(2\nu+1)$&$\frac{1}{3}(\nu-1)$
\end{tabular}
\end{table}
For the $K^{*0}(892)$ production we have the $d\bar{s}$ matrix element of the matrix $M$,
and the
$s\bar{d}$ element for hadronization, which means we need now the $(3,2)$ element of the matrix
$\textbf{V}\cdot \textbf{V}$ upon hadronization and, as a consequence, considering the states
\begin{eqnarray}
 |\bar{K}^*\rho,I=1/2,I_3=1/2\rangle&=&\sqrt{\frac{2}{3}}|\rho^+K^{*-}\rangle-\frac{1}{\sqrt{3}}|\rho^0 \bar{K}^{*0}\rangle,\\
|\bar{K}^*\omega,I=1/2,I_3=1/2\rangle&=&|\bar{K}^{*0}\omega\rangle,\\
|\bar{K}^*\phi,I=1/2,I_3=1/2\rangle&=&|\bar{K}^{*0}\phi\rangle,
\end{eqnarray}
which are the building blocks of the $\bar{K}_2^*(1430)$ resonance, we obtain the weights $W^{(K^{*\,0})}_i$ for the $J/\psi\rightarrow K^{*0}V'V'$ transition (Table \ref{weight3}).

\begin{table}[h!]
\caption{Weights for the process $J/\psi\rightarrow K^{*\,0}V'V'$ with $V^\prime V^\prime$ in isospin $I=1/2$, $I_{3}=1/2$. }\label{weight3}
\vspace{0.3cm}
\centering
\begin{tabular}{c|ccc}
\hline\hline\\
&$\rho\bar{K}^*$&$ \bar{K}^*\omega$&$\bar{K}^*\phi$\\
\hline
&&&\\
$W^{(K^{*\,0})}$&$\sqrt{\frac{3}{2}}$&$\sqrt{\frac{1}{2}}$&1
\end{tabular}
\end{table}

\subsection{Dynamical generation of the tensor resonances}
\begin{figure}[h!]
\includegraphics[width=\textwidth]{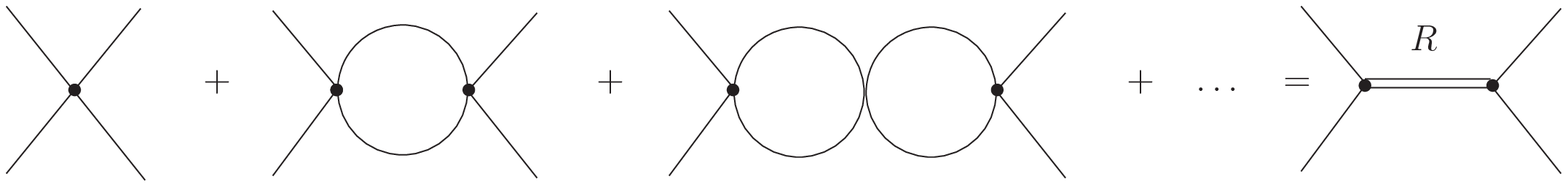}
\caption{Diagrammatical representation of  Eq. (\ref{T}).}\label{BS}
\end{figure}

In Ref.~\cite{gengvec} the resonances were obtained by solving the Bethe Salpeter
equation in coupled channels to obtain the scattering matrix

\begin{equation}
T=[1-\tilde{V}G]^{-1}\tilde{V}\label{T}
\end{equation}
where $\tilde{V}$ is the transition potential from $V^\prime V^\prime\rightarrow V^\prime V^\prime$
and $G$ is the loop function of two mesons given in \cite{gengvec}. Searching for poles of the amplitudes provides the mass and width\footnote{One should note that the total width of these states is partly due to decay into two pseudoscalar mesons, non negligible mostly because of the large phase space for this decay, but the two pseudoscalar meson channels play a minor role in the generation of these resonances \cite{gengvec}.} of the states as well as the coupling, $g_{i}$, of the resonances to the different $V^\prime V^\prime$ channels, since close to the pole the amplitude behaves as
\begin{equation}
T_{ij}=\frac{g_{i}g_{j}}{z-z_{R}}.
\end{equation}
Diagrammatically one is summing the diagrams of Fig. (\ref{BS}).

\begin{figure}[h!]
\includegraphics[width=\textwidth]{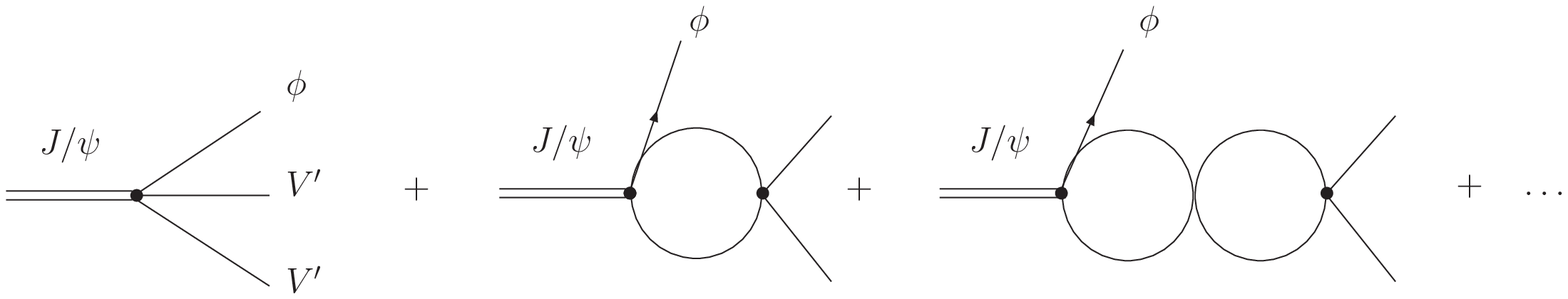}
\caption{Production mechanism  of $\phi$ and two interacting vector mesons.}\label{vertexJ}
\end{figure}

\begin{figure}[h!]
\centering
\includegraphics[scale=0.6]{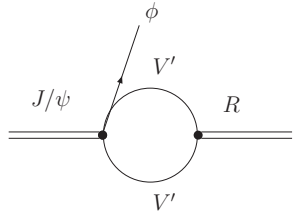}
\caption{Selection of diagrams of Fig. \ref{vertexJ} that go into resonance formation omitting its final coupling to $V^\prime V^\prime$.}\label{JRes}
\end{figure}

Hence the $V^\prime V^\prime$ production, upon consideration of the interaction of $V^\prime V^\prime$, proceeds as shown in Fig. \ref{vertexJ},
the resonance contribution in $J/\psi \to \phi R$ can be depicted diagrammatically as in Fig. \ref{JRes},
which involves the weights $J/\psi\to\phi V^\prime V^\prime$, the $G$ functions, or loop function of two meson propagators, and the couplings $g_{i}$ of the resonance to the different $V^\prime V^\prime$ pairs. The couplings $g_{i}$ and $G_{i}$ functions for the different channels and resonances are obtained in \cite{gengvec} and we reproduce them in Table \ref{coupG}. We take the real part of the $G$ functions once we realize that keeping the small imaginary part only induces minor corrections. The estimation of the theoretical errors has been done by
changing some parameters of the theory according to the following rules:

\begin{itemize}

\item The value of $g_v=\frac{2 M_V}{f}$. In Ref.~\cite{gengvec}, we have used
the mass of the rho meson and the decay constant of the pion to derive the
value for $g_v$. An alternative is to use the average mass of the vector nonet
and the average decay constant of the pion octet. In this way, one obtains a
smaller value: $g_v'\approx0.94 g_v$. To estimate the uncertainties related to
this parameter we use a smaller value for this coupling, $g'_v=0.9g_v$,
and change the values of the subtraction constants  (which appear in the
dimensional formulas for G in \cite{gengvec}) in the $K^*\bar{K}^*$  and
$\rho\rho$ channels to refit the experimental $f_2(1270)$ and $f'_2(1525)$
masses (keeping the values of the subtraction constants of other channels
unchanged).

\item The values of the subtraction constants. One can of course slightly
change the values of the subtraction constants. Since the masses of the
$f_2(1270)$ and $f'_2(1525)$ are rather accurately determined, this implies
that our modification of the values of the $\rho\rho$, $K^*\bar{K}^*$
subtraction constants must be small and the modification of those of the less
important channels can be relatively large. More specifically, we can roughly
change the values of the  $\rho\rho$ and $K^*\bar{K}^*$ subtraction constants
by 2\%, while changing those of the $\omega\omega$ and $\phi\phi$ by 20\% and
10\%, respectively, to keep the masses of these two resonances within experimental errors.
\end{itemize}

In a similar way, one can obtain the uncertainties inherent in the $g_{i}$ and $G_{i}$ of the $\bar{K}^{*\,0}_{2}(1430)$.

In principle in Fig. \ref{JRes} one should symmetrize the amplitudes for $\phi$ exchange in the case that $V^\prime V^\prime$ correspond to the $\phi\phi$ component. In practice the external $\phi$ and those in the loop ($V^\prime$) are kinematically very different and the symmetrization of the external $\phi$ with one $V^\prime$ is of minor relevance. This, together with the fact that the $\phi\phi$ components are small in the present case, as seen in Table \ref{coupG}, justifies ignoring this issue. The same can be said for the case of an external $\omega$.

\begin{table}[h]
\caption{Couplings and values of the loop functions for the different channels at the resonance energy.}
\vspace{0.3cm}
\centering
\begin{small}
\begin{tabular}{l c| c ccccccc}
\hline\hline
&&&&&&&&&\\
 &&  &$\rho\rho$&$K^* \bar{K}^*$&$\omega\omega$&$\phi\phi$&$\rho\bar{K}^*$&$\omega\bar{K}^{*\,0}$&$\phi\bar{K}^{*\,0}$\\
[0.1ex]
\hline
 & &$g_{i} (\textrm{MeV})$ &10551 & $4771$ & $-503$ & $-771$&0&0&0   \\[1ex]
 & & error $g_{i} (\%) $
&  $4$ & $3$ & $22$ & 22&0&0&0  \\[1.ex]
\raisebox{1.5ex}{}  &\raisebox{1.5ex}{$f_{2}(1270)$} &$G_{i} (\times 10^{-3})$
&  $-4.74$ & $-3.00$ & $-4.97$ & 0.475&0&0&0  \\[1.ex]
& &error $G_{i} (\%)$
&  $10$ & $29$ & $42$ & 220&0&0&0  \\[1.ex]

 \hline
& &$g_{i} (\textrm{MeV})$ & $-2611$ & $9692$ & $-2707$ & $-4611$&0&0&0   \\[1ex]
 & & error $g_{i} (\%) $
&  $12$ & $6$ & $2$ & 2&0&0&0  \\[1.ex]

\raisebox{1.5ex}{} & \raisebox{1.5ex}{$f^\prime_{2}(1525)$}& $G_{i} (\times 10^{-3})$
&$-8.67$ & $-4.98$ & $-9.63$ & $-0.710$&0&0&0 \\[1ex]
& &error $G_{i} (\%)$
&  $6$ & $17$ & $19$ & 141&0&0&0  \\[1.ex]

\hline
& &$g_{i} (\textrm{MeV})$ & 0 & $0$ & 0 & 0&10613&2273&$-2906$   \\[1ex]
 & & error $g_{i} (\%) $
&  $0$ & $0$ & $0$ & 0&3&5&5  \\[1.ex]
\raisebox{1.5ex}{} & \raisebox{1.5ex}{${\bar{K}}_{2}^{*\,0}(1430)$}& $G_{i} (\times 10^{-3})$
&0& $0$ & 0 & 0&$-6.41$&$-5.94$&$-2.70$ \\[1ex]
& &error $G_{i} (\%)$
&  $0$ & $0$ & $0$ & 0&12&19&43  \\[1.ex]

\hline
\end{tabular}
\end{small}
\label{coupG}
\end{table}

The final transition matrix for $J/\psi\to\phi(\omega,{K^*}^0)R$ is given by
\begin{equation}
t_{J/\psi\to\phi R}=\sum_{j}W_{j}^{(\phi)}G_{j}g_{j}\label{tJ}
\end{equation}
and the same for $J/\psi\to\omega({K^*}^0)R$, with the weights given by
Tables \ref{coef2}, \ref{weight2}, and \ref{weight3} and the $G$ functions and couplings given in Table \ref{coupG}.

\section{Results}
The $J/\psi$ decay widths are given by
\begin{equation}
\Gamma=\frac{1}{8\pi}\frac{1}{M^2_{J/\psi}}|t|^2q
\end{equation}
with $q$ the momentum of the $\phi (\omega, K^{*\,0})$ in the $J/\psi$ rest frame. The experimental data that we use are given in Table \ref{data} .
\begin{table}[h!]
\caption{Experimental branching ratios.}
\begin{center}
\begin{tabular}{lcc}
\hline\hline
&Obtained from&$BR[\times 10^{-3}]$\\
\hline
$J/\psi\to\omega f_{2}(1270)$&PDG&4.3$\pm$0.6\\
$J/\psi\to\phi f_{2}(1270)$&BES(\cite{ablikim1})&0.27$\pm$0.06\\
$J/\psi\to\omega f^\prime_{2}(1525)$&PDG&$<$0.2\\
$J/\psi\to\phi f^\prime_{2}(1525)$&BES(\cite{ablikim1})&0.82$\pm$0.12\\
$J/\psi\to K^{*\,0} \bar{K}^{*\,0}_{2}(1430)$&PDG&3.0$\pm$0.3\\
\hline
\end{tabular}
\label{data}
\end{center}
\end{table}

For the branching ratios of $J/\psi\to\omega f_{2}(1270)$, $J/\psi\to\omega f^\prime_{2}(1525)$ and
$J/\psi\to K^{*\,0} \bar{K}^{*\,0}_{2}(1430)$ we have taken the well established PDG data \cite{pdg}. For the $J/\psi\to\phi f_{2}(1270)$ and $J/\psi\to\phi f^\prime_{2}(1525)$ branching ratios we have instead taken the most recent data from the BES collaboration \cite{ablikim1}. The data from the DM2 collaboration \cite{falvard} depend upon assumptions of possible interference with other resonances, but in all cases the ratio of these two rates, which is what we need for our work, is similar to those obtained by BES \cite{ablikim1}. The ratios obtained from \cite{gidal}, after comparison of different data, are of the same order of magnitude but with larger uncertainties. For our fit we have taken the results from the most recent data of BES \cite{ablikim1}. For the $J/\psi\to\omega f^\prime_{2}(1525)$ branching ratio the PDG quotes a value BR $< 0.2\times 10^{-3}$. Once again BES \cite{ablikim} has recent data for that, from where one induces that the branching ratio is around $0.2\times 10^{-3}$  with uncertainties that could be as large as $50\%$, which is the error assumed in Table \ref{res} when we perform a fit to the data evaluating $\chi^2$. Assuming this large error guarantees a smaller weight of this datum, not precisely known experimentally, in the global fit.

In order to get the different decay rates we fit the only parameter we have, $\nu$.
Since we took $T^{(8,8)}\equiv 1$, thus fixing an arbitrary normalization, we can only obtain ratios between the partial decay rates. We choose the ratios
\begin{equation}
R_{1}\equiv\frac{\Gamma_{J/\psi\to\phi f_{2}(1270)}}{\Gamma_{J/\psi\to\phi f^\prime_{2}(1525)}},\quad
R_{2}\equiv\frac{\Gamma_{J/\psi\to\omega f_{2}(1270)}}{\Gamma_{J/\psi\to\omega f^\prime_{2}(1525)}},\label{ratios1}
\end{equation}
\begin{equation}
R_{3}\equiv\frac{\Gamma_{J/\psi\to\omega f_{2}(1270)}}{\Gamma_{J/\psi\to\phi f_{2}(1270)}},\quad
R_{4}\equiv\frac{\Gamma_{J/\psi\to K^{*\,0} \bar{K}^{*\,0}_{2}(1430)}}{\Gamma_{J/\psi\to\omega f_{2}(1270)}}.\label{ratios2}
\end{equation}
Upon minimization of the $\chi^2$ function we obtain an optimal solution with
the value of the parameter $\nu=1.45$. The theoretical results are given in Table \ref{res} and compared to the experimental brackets.  The errors in Table \ref{res} have been estimated taking the errors of Table \ref{coupG} and summing in quadrature the errors of the terms in Eq. (\ref{tJ}).

\begin{table}[h]
\caption{Comparison between the experimental and the theoretical results.}
\begin{center}
\begin{tabular}{ccc}
\hline\hline
&Experiment&Theory\\
\hline
\\
$R_{1}$&0.22 - 0.47 $(0.33^{+0.14}_{-0.11})$&0.13 - 0.61 $(0.28^{+0.33}_{-0.15})$\\[1.5ex]
$R_{2}$&12.33 - 49.00 $(21.50^{+27.50}_{-9.17})$&2.92 - 13.58 $(5.88^{+7.70}_{-2.96})$\\[1.5ex]
$R_{3}$&11.21 - 23.08 $(15.85^{+7.23}_{-4.65})$&6.18 - 19.15 $(10.63^{+8.52}_{-4.45})$\\[1.5ex]
$R_{4}$&0.55 - 0.89 $(0.70^{+0.19}_{-0.15})$&0.83 - 2.10 $(1.33^{+0.77}_{-0.50})$\\[1.5ex]
\hline
\end{tabular}
\label{res}
\end{center}
\end{table}

The overall agreement with data is reasonable. We obtain four independent ratios with just one parameter. On the other hand this parameter can be related to $\lambda_{\phi}$ of \cite{ulfjose} as done in \cite{palochiang} through
\begin{equation}
\lambda_{\phi}=\sqrt{2}\Bigg (\frac{\nu-1}{\nu+2}\Bigg)
\end{equation}
which provides a value of $\lambda_{\phi}=0.18$, very close to the one obtained
 in \cite{ulfjose,palochiang}, $\lambda_{\phi}=0.13-0.20$. This relationship  is readily obtained dividing the two terms in the $\phi$ production column of Table \ref{coef}. Although we have different physics than in
 \cite{ulfjose,palochiang} since we have the production of pairs of vectors
 rather than pseudoscalars, and we have also tensor states rather than scalars,
 it is gratifying to see that the value of $\lambda_{\phi}$, which is a measure
 of the subdominant mechanism of Fig. \ref{ccbar}b in $J/\psi\to \phi V^\prime V^\prime$, is a small number, comparable in
 size and sign to the one obtained in \cite{ulfjose,palochiang}.

\begin{figure}
\includegraphics[width=\textwidth]{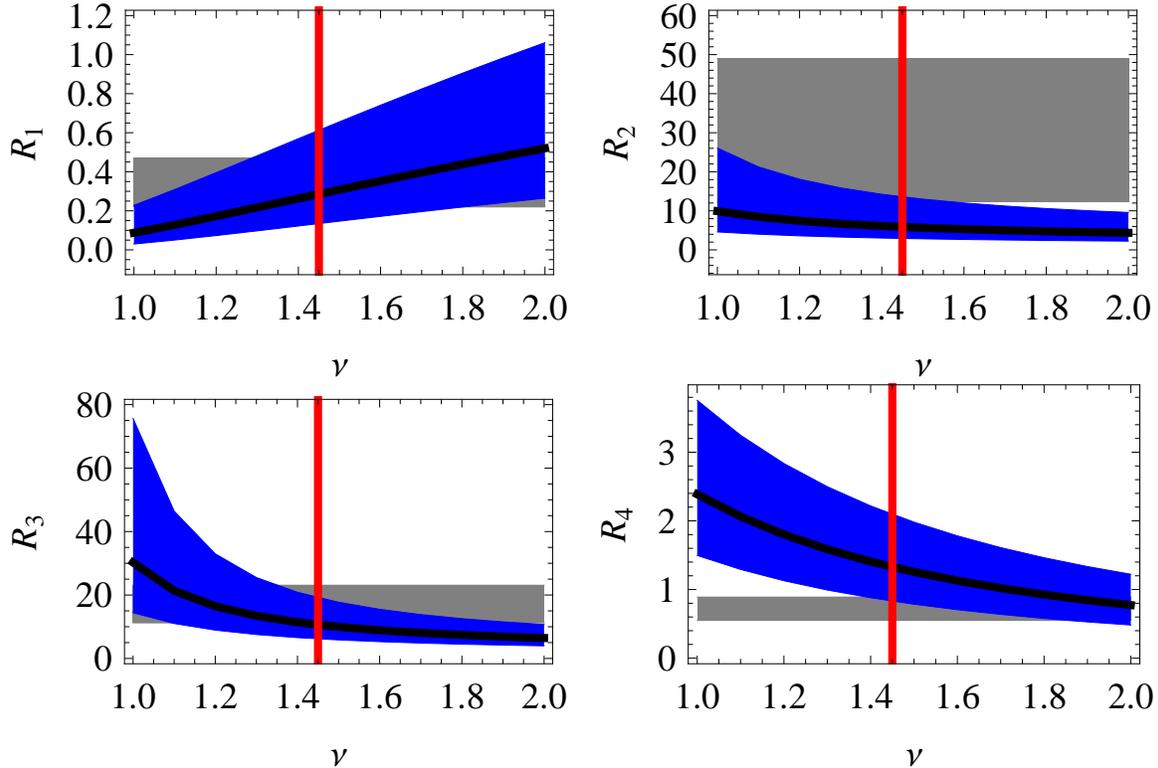}
\caption{Comparison with data (gray shaded areas) of $R_{1}$, $R_{2}$, $R_{3}$, and $R_{4}$ calculated with Eqs. (\ref{ratios1}, \ref{ratios2}). The vertical solid red line indicate $\nu=1.45$ and the blue shaded areas indicate the range of values for $R_{1}$, $R_{2}$, $R_{3}$, and $R_{4}$ as a function of $\nu$.} \label{new1}
\end{figure}

We have shown the results obtained for the value of $\nu$ that provides the best fit to the data. It is also interesting to show how the theoretical results and their uncertainties change with the parameter $\nu$. This is shown in Fig. \ref{new1} (note different scales for the different ratios). What we observe there is that with large values of $\nu$ the agreement for $R_{2}$ is spoiled, while for smaller values of $\nu$ the ratio $R_{4}$ becomes incompatible with the data.

\begin{figure}
\includegraphics[width=\textwidth]{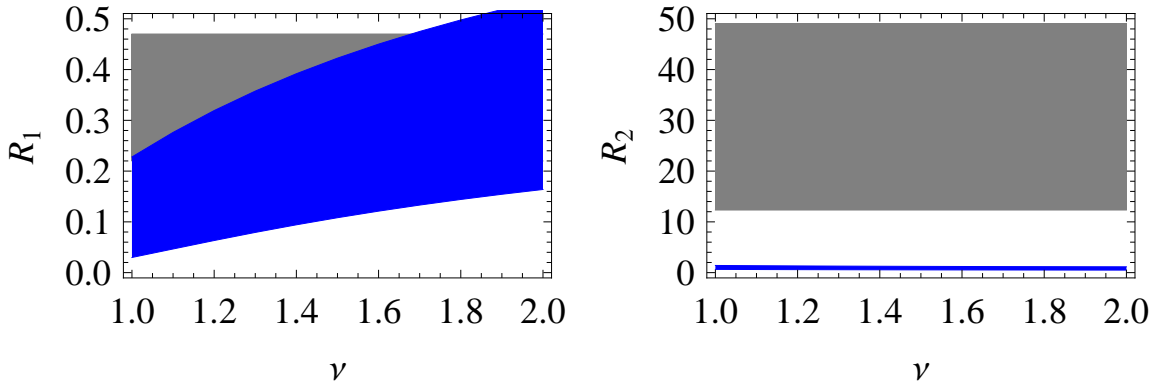}
\caption{Comparison with data (gray shaded areas) of $R_{1}$ and $R_{2}$ calculated with Eqs. (\ref{ratios1}) by changing the sign of the $\rho \rho$  coupling to the $f^\prime_{2}(1525)$  in Table \ref{coupG} . The blue shaded areas indicate the range of values for $R_{1}$ and $R_{2}$ as a function of the $\nu$ parameter.}
\label{new2}
\end{figure}

The success in the description of the experimental data is by no means trivial
and can be traced back to the particular couplings of the resonance to the
$V^\prime V^\prime$ states. Note that the important couplings to $\rho\rho$ and
$K^*\bar{K}^*$ have the same relative sign for the $f_{2}(1270)$ and opposite relative sign for the $f^{\prime}_{2}(1525)$. This feature is essential to the success of the results.
 Should all the couplings have the same sign it would have been impossible to
 get any reasonable fit to the data. As an example, in Fig. \ref{new2}, we give the results that we would obtain if we change the relative sign of the $\rho \rho$ and $K\bar{K}^*$ couplings to the $f^\prime_{2}(1525)$ (we take now $g=2611$ MeV for the $\rho\rho$ coupling in Table \ref{coupG}). While the ratio $R_{1}$ would still be compatible with the data, the ratio $R_{2}$ is now within 0.50-1.41, an order of magnitude smaller than the lower value of the experimental interval.

\section{Conclusions}
 We have carried out an evaluation of the rates of the
 $J/\psi \to \phi (\omega) f_2(1270)$,
$J/\psi \to \phi (\omega) f'_2(1525)$ and $J/\psi \to K^{*0}(892) \bar{K}^{*\,0}_2(1430)$
decay modes of $J/\psi$.  The basic element to perform the calculations was the
assumption that these tensor states are all dynamically generated from the
vector-vector interaction provided by the hidden gauge formalism, upon
unitarization in coupled channels. In a previous work, where these states were
obtained with this formalism, the couplings of the resonances to the pairs of
vector-vector building blocks  were evaluated and we have used these results
here. Another ingredient necessary in the present work was the evaluation of the
primary weights for $J/\psi$ decay into $\phi, \omega, K^{*0}(892)$ plus a pair
of the vector building blocks of the resonances. This was accomplished by taking
the  $J/\psi$  as a singlet of SU(3) and
considering ideal mixing for the $\phi$ and $\omega$ states.  The results obtained for
the four independent ratios of the rates are in reasonable agreement with experiment. The
success of this comparison with ratios, which vary in one or two orders of magnitude
from one to another, is by no means trivial. It sensibly depends upon the strength
and the signs of the couplings of the resonances to the different vector-vector
channels. Our study shows that the theoretical results obtained from the idea of these tensor mesons being dynamically generated from the vector-vector interaction are compatible with present data. However, we should keep in mind that the experimental uncertainties are still large, and further accuracy on the data would be most welcome to provide a more stringent test on the theoretical approach.

\section*{Acknowledgments}
We would like to thank C. Hanhart for useful discussions and suggestions.
This work is partly supported by DGICYT contract number
FIS2006-03438, the National Natural Science Foundation of China (No.
10675058, No. 10875133, and No. 10775012) and the Scientific Research Foundation of Liaoning
Education Department (No. 20060490 and No. 2009T055). We acknowledge the support of
the European Community-Research Infrastructure Integrating Activity
``Study of Strongly Interacting Matter" (acronym HadronPhysics2.
Grant Agreement n. 227431) under the Seventh Framework Programme of
EU. A. M. T is supported by a FPU grant of the Ministerio de Ciencia
e Innovaci\'on. L.S.G. acknowledges support from the MICINN in the
Program ÒJuan de la Cierva.Ó


\begin{thebibliography}{99}

\bibitem{ulfjose}
 U.~G.~Meissner and J.~A.~Oller,
 Nucl.\ Phys.\  A {\bf 679}, 671 (2001).

\bibitem{palochiang}
 L.~Roca, J.~E.~Palomar, E.~Oset and H.~C.~Chiang,
 Nucl.\ Phys.\  A {\bf 744}, 127 (2004).

\bibitem{lahde}
 T.~A.~Lahde and U.~G.~Meissner,
 Phys.\ Rev.\  D {\bf 74}, 034021 (2006).
\bibitem{liu}
  B.~Liu, M.~Buescher, F.~K.~Guo, C.~Hanhart and U.~G.~Meissner,
  arXiv:0901.1185 [hep-ph].

\bibitem{npa}
 J.~A.~Oller and E.~Oset,
 Nucl.\ Phys.\  A {\bf 620}, 438 (1997)
 [Erratum-ibid.\  A {\bf 652}, 407 (1999)]
 .

\bibitem{ramonet}
 J.~A.~Oller, E.~Oset and J.~R.~Pelaez,
 Phys.\ Rev.\  D {\bf 59}, 074001 (1999)
 [Erratum-ibid.\  D {\bf 60}, 099906 (1999\ ERRAT,D75,099903.2007)]
 .

\bibitem{kaiser}
 N.~Kaiser,
 Eur.\ Phys.\ J.\  A {\bf 3}, 307 (1998).

\bibitem{markushin}
 V.~E.~Markushin,
 Eur.\ Phys.\ J.\  A {\bf 8}, 389 (2000)
 .

\bibitem{rios}
 J.~R.~Pelaez and G.~Rios,
 Phys.\ Rev.\ Lett.\  {\bf 97}, 242002 (2006)
 .

\bibitem{arriola}
 J.~Nieves and E.~Ruiz Arriola,
 Nucl.\ Phys.\  A {\bf 679}, 57 (2000)
 .

\bibitem{juanlarge}
 J.~Nieves and E.~R.~Arriola,
 arXiv:0904.4344 [hep-ph].

\bibitem{raquel}
 R.~Molina, D.~Nicmorus and E.~Oset,
 Phys.\ Rev.\  D {\bf 78}, 114018 (2008).





\bibitem{hidden1}
 M.~Bando, T.~Kugo, S.~Uehara, K.~Yamawaki and T.~Yanagida,
 Phys.\ Rev.\ Lett.\  {\bf 54}, 1215 (1985).



\bibitem{hidden2}
 M.~Bando, T.~Kugo and K.~Yamawaki,
 Phys.\ Rept.\  {\bf 164}, 217 (1988).

\bibitem{hidden3}
 M.~Harada and K.~Yamawaki,
 Phys.\ Rept.\  {\bf 381}, 1 (2003)
 .


\bibitem{gengvec}
 L.~S.~Geng and E.~Oset,
 Phys.\ Rev.\ D {\bf 79}, 074009 (2009).
L.~S.~Geng, E.~Oset, R.~Molina and D.~Nicmorus,
arXiv:0905.0419 [hep-ph].



\bibitem{pdg}
 C.~Amsler {\it et al.}  [Particle Data Group],
 Phys.\ Lett.\  B {\bf 667}, 1 (2008).

\bibitem{ablikim1}
  M.~Ablikim {\it et al.}  [BES Collaboration],
  Phys.\ Lett.\  B {\bf 607}, 243 (2005)
  .

 \bibitem{falvard}
  A.~Falvard {\it et al.}  [DM2 Collaboration],
  Phys.\ Rev.\  D {\bf 38}, 2706 (1988).
\bibitem{gidal}
  G.~Gidal {\it et al.},
  Phys.\ Lett.\  B {\bf 107}, 153 (1981).

\bibitem{ablikim}
  M.~Ablikim {\it et al.}  [BES Collaboration],
  Phys.\ Lett.\  B {\bf 603}, 138 (2004)
  .






\end{thebibliography}
\end{document}